\documentclass[aps,pra, reprint,superscriptaddress,a4paper]{revtex4-1} 
\usepackage{amsmath,amssymb} 
\usepackage[utf8]{inputenc} 
\usepackage{graphicx} 
\usepackage{xcolor} 
\usepackage{hyperref}
\hypersetup{colorlinks=true,citecolor={blue},linkcolor={blue},urlcolor={blue}} 
\usepackage{braket}
\usepackage{multibib}
\usepackage{bm}

\usepackage[separate-uncertainty=true,multi-part-units=single,group-separator={,}]{siunitx} 
\DeclareSIUnit[per-mode=symbol,per-symbol=p]{\ueV}{\micro\electronvolt}
\DeclareSIUnit[per-mode=symbol,per-symbol=p]{\um}{\micro\meter}

\newcommand{\mi}{\mathrm{i}}

\newcommand{\ve}{\varepsilon}
\newcommand{\half}{\textstyle{\frac{1}{2}}}
\newcommand{\ihalf}{\textstyle{\frac{\mathrm{i}}{2}}}

\graphicspath{{./figs/}}
\usepackage{float}
\begin{document}

\title{Tunable magnetic alignment between trapped exciton-polariton condensates}

\author{H. Ohadi}
\email{ho278@cam.ac.uk}
\author{Y. del Valle-Inclan Redondo}
\author{A. Dreismann}
\affiliation{Department of Physics, Cavendish Laboratory, University of Cambridge, Cambridge CB3 0HE, United Kingdom}
\author{Y. G. Rubo} \affiliation{Instituto de Energías Renovables, Universidad Nacional Autónoma de México, Temixco, Morelos, 62580, Mexico} 
\author{F. Pinsker} \affiliation{Clarendon Laboratory, University of Oxford, Parks Road, Oxford OX1 3PU, United Kingdom} 
\author{S. I. Tsintzos} \affiliation{ Foundation for Research and Technology–Hellas, Institute of Electronic Structure and Laser, 71110 Heraklion, Crete, Greece }
\author{Z. Hatzopoulos} \affiliation{ Foundation for Research and Technology–Hellas, Institute of Electronic Structure and Laser, 71110 Heraklion, Crete, Greece } \affiliation{ Department of Physics, University of Crete, 71003 Heraklion, Crete, Greece }
\author{P. G. Savvidis}
\affiliation{ Foundation for Research and Technology–Hellas, Institute of Electronic Structure and Laser, 71110 Heraklion, Crete, Greece } 
\affiliation{ Department of Materials Science and Technology, University of Crete, 71003 Heraklion, Crete, Greece }
\author{J. J. Baumberg}
\email{jjb12@cam.ac.uk}
\affiliation{Department of Physics, Cavendish Laboratory, University of Cambridge, Cambridge CB3 0HE, United Kingdom}

\begin{abstract}
Tunable spin correlations are found to arise between two neighboring trapped
exciton-polariton condensates which spin-polarize spontaneously. We observe a
crossover from an antiferromagnetic- to a ferromagnetic pair state by reducing
the coupling barrier in real-time using control of the imprinted pattern of pump light.
Fast optical switching of both condensates is then achieved by resonantly but weakly
triggering only a single condensate. These effects can be explained
as the competition between spin bifurcations and spin-preserving Josephson
coupling between the two condensates, and open the way to polariton Bose-Hubbard ladders.
\end{abstract}

\maketitle

The development of spin-charge lattice models for understanding strongly-correlated states
of matter is a successful theme of modern quantum physics. This has driven the desire to model and probe 
 complex condensed matter phenomena using
highly-controlled systems, such as ultracold atoms~\cite{bloch_many-body_2008},
photons~\cite{szameit_discrete_2010,hafezi_imaging_2013}, or superconducting
junctions~\cite{you_atomic_2011}. Exciton-polariton (polariton) lattices have
emerged as an alternative system~\cite{carusotto_quantum_2013,
    jacqmin_direct_2014} with unique properties. Due to their strongly
dissipative and nonlinear nature, many-body polariton gases can reach steady
states which are remarkably different from their equilibrium
case~\cite{le_boite_steady-state_2013}. Moreover, they have peculiar spin
properties~\cite{leyder_observation_2007,lagoudakis_observation_2009,sala_spin-orbit_2015}
and exhibit spontaneous magnetization (emitting circularly polarized light)
above a critical bifurcation threshold~\cite{ohadi_spontaneous_2015}, analogous
to the weak lasing regime~\cite{aleiner_radiative_2012}. In this Letter we study the basic
building block of a polariton spin lattice: two optically trapped
spin-polarized condensates which are tunably coupled. We demonstrate that
trapped out-of-equilibrium polariton condensates can exhibit Ising-like behavior
related to spin bifurcations. The two condensate system investigated here is
shown to correspond to one plaquette of a bosonic ladder~\cite{atala_observation_2014}, and allows
demonstration of a crossover in the competition between Josephson coupling and
spin bifurcation. These features have not been seen in any other system to date.

Polariton condensates are coherent many-body
states~\cite{deng_condensation_2002, kasprzak_bose-einstein_2006,
    balili_bose-einstein_2007, baumberg_spontaneous_2008}, which can be confined
in potentials~\cite{wertz_spontaneous_2010, tosi_sculpting_2012,
    cristofolini_optical_2013} and interact with each other via Josephson
junctions~\cite{lai_coherent_2007, lagoudakis_coherent_2010,
    tosi_geometrically_2012, abbarchi_macroscopic_2013}.  For a pair of
interacting trapped spin-polarized condensates, their polarization states are
expected to couple.  However, the driven-dissipative and nonlinear nature of
polariton condensates makes the underlying coupling mechanism considerably
richer than that in the conventional Ising case, leading to exotic forms of
magnetism where the orientation and strength of coupling is not determined by
the sign of the interaction. 

\begin{figure}[t!] \centering \includegraphics[width=0.45 \textwidth]{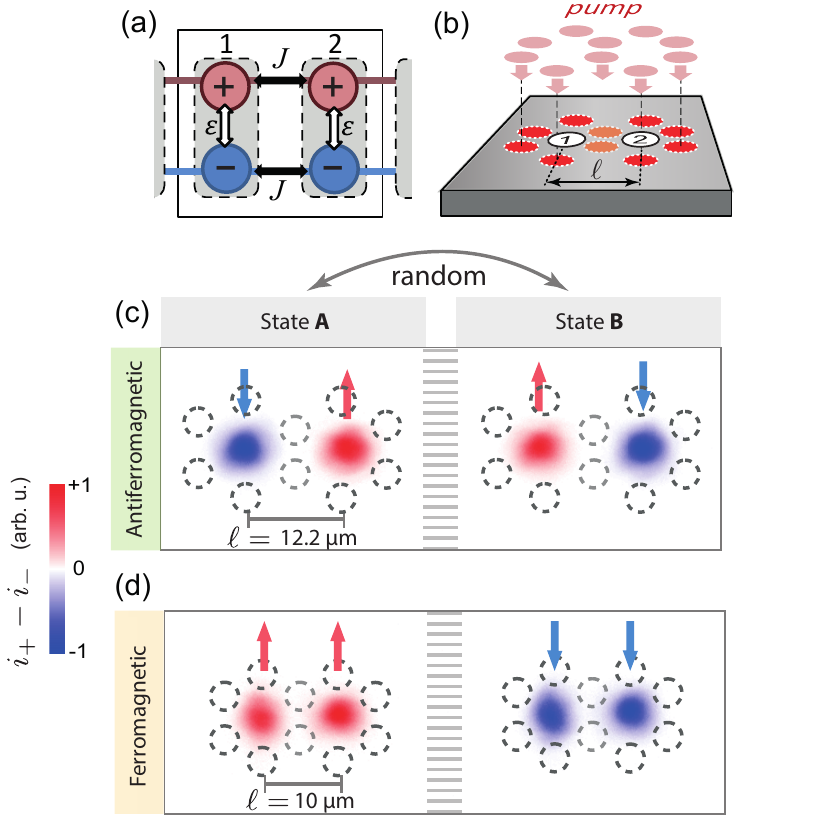}
    \caption{(a) Coupling between condensates 1,2 controlled by Josephson tunneling $J$, spin-coupling ($+ \leftrightarrow -$) within condensates controlled by energy splitting $\varepsilon$. (b) Schematic double condensate trap from 10 pump beams. (c,d)
		Experimental spin states seen for (c) AF- and (d) F-coupled condensates. In each case two possible states exist, with actual state chosen randomly upon each realization.} \label{fig:DoubleHex} \end{figure}

We achieve tuning between ferromagnetic (F) or antiferromagnetic (AF) coupling
by directly modulating the tunneling barrier, adjusting either the height of the barrier or the separation between the condensates.  We show how optical
switching of the spin state of one condensate results in fast switching of the
state of the neighboring condensate. Our result is a key step towards using
trapped polariton condensates for the realization of interacting bosons in a
driven-dissipative spinor Bose-Hubbard model~\cite{le_boite_steady-state_2013,
    carusotto_quantum_2013}.

The Bose-Hubbard model with polaritons can be studied in other systems where,
for example, the sample is etched to form micro-pillar
arrays~\cite{abbarchi_macroscopic_2013, jacqmin_direct_2014,
    sala_spin-orbit_2015}, metal films are deposited on the surface of the
cavity~\cite{lai_coherent_2007,kim_gaas_2008, kim_dynamical_2011}, or surface
acoustic waves are applied~\cite{cerda_mendez_polariton_2010}. While the confinement potential is
then separate from the condensate gain, the trapped exciton-polaritons studied in the
present letter have the advantage that the confinement, even at a single site
level, is versatile and can be adjusted on the fly. Particularly for larger arrays of condensates, this capability to tune the different barriers is vital. Moreover, the interaction with the reservoir particles is reduced, crucial for spin stability.

Exciton-polaritons (polaritons) are quasiparticles formed by the strong coupling
of excitons in semiconductor quantum wells with photons in the microcavity in
which they are embedded~\cite{kavokin_microcavities_2007}.  We create optically
trapped polariton condensates~\cite{cristofolini_optical_2013,
    askitopoulos_polariton_2013, dreismann_coupled_2014} by nonresonant
linearly-polarized continuous wave (CW) excitation of a membrane
microcavity~(for details see SI.~\ref{si:expmeth}). Driven by their repulsive
excitonic interactions, polaritons travel away from the pump region and feed the
zero-momentum ground state at the center of the optical trap.  Once the density
exceeds the condensation threshold, a macroscopically coherent condensate 
forms in the trap center. Polaritons in quantum-well microcavities have two
$\pm 1$ (spin up or down) projections of their total angular momentum along the
structure growth axis, which correspond to right- and
left-circularly polarized photons emitted by the cavity (of intensities
$i_\pm$).

Trapped polariton condensates spontaneously exhibit a high degree of circular
polarization, or magnetization $M=s_z=(i_+-i_-)/(i_++i_-)$, above a critical spin-bifurcation threshold as a result of
energy and dissipation splitting of their linear
polarizations~\cite{ohadi_spontaneous_2015}. We operate above this
spin-bifurcation threshold, which means that the trapped condensate is
spin-polarized in either spin-up $\ket{\uparrow}$ or spin-down
$\ket{\downarrow}$ states. These spin-polarized
condensates emit nearly circularly polarized ($\vert M \vert>$85\%) light, which can be
measured using conventional polarimetry. In each realization we
excite the sample for $\SI{200}{\us}$ and measure
the condensate $M_n(t)$, where $n \in \{1,2\}$ denotes the left and right condensate (see SI.~\ref{si:realspace}). The optical
excitation is patterned using a spatial light modulator (SLM) into the shape of
a double-hexagon as shown by dashed circles in Fig.~\ref{fig:DoubleHex}(b-d) such
that a spin-polarized condensate is formed at the center of each hexagon. The
middle `barrier' pump spots [orange in Fig.~\ref{fig:DoubleHex}(b)] between the two traps are weaker than the outer spots
(intensity ratio $\simeq 75\%$) to allow inter-trap tunneling of polaritons.

We can spatially squeeze or stretch the traps and change the condensates
separation without changing the barrier pump intensity. To achieve this the
barrier pump spots of the double-hexagon trap are fixed while shifting the location of
the other spots. When the separation of the condensates maxima is greater than
$\ell_c=\SI{13.6}{\um}$, the two condensates independently pick a spin-up or spin-down
state. However, when this is decreased to $\ell=0.90\ell_c$,
we observe AF coupling, where the condensates spontaneously collapse into either
$\ket{\uparrow\downarrow}$ or $\ket{\downarrow \uparrow}$ states in each
realization [Fig.~\ref{fig:DoubleHex}(c), see also
SI.~\ref{si:realspace}]. Further decreasing the separation to
$0.74\ell_c$, we observe F coupling where the condensates pick either of
$\ket{\uparrow \uparrow}$ or $\ket{\downarrow \downarrow}$ states randomly in
each realization [Fig.~\ref{fig:DoubleHex}(d)]. As in the case of a
single condensate~\cite{ohadi_spontaneous_2015}, these states remain stable for
many seconds at \SI{5}{\kelvin}, and do not depend on the position on the
sample, the geometrical pattern of the pump spots or the power above the
spin-bifurcation threshold.

\begin{figure} \centering \includegraphics[width=0.5
    \textwidth]{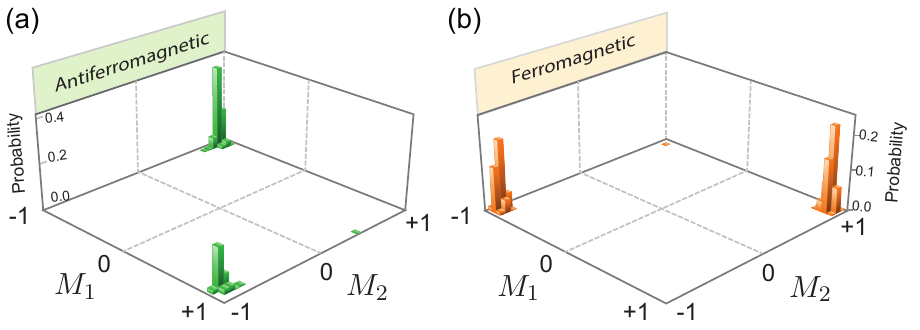}
    \caption{2D histogram of correlations in measured double condensate spins $(M_1,M_2)$ over \num{1000} realizations in
        the (a) antiferromagnetic and (b) ferromagnetic coupling regimes.} \label{fig:Correlations} \end{figure}

For each set of trapping conditions, \num{1000} realizations are created and for
each we measure $M$ for the two condensates to perform a statistical analysis of
the condensate-pair spin correlation. Each polarization-resolved realization is
recorded for \SI{200}{\us} by a camera, allowing us to  map the 2D histogram of
$(M_1,M_2)$ and resolving F and AF situations (Fig.~\ref{fig:Correlations}).
Absolute correlations of $\vert C \vert>0.99$ are found for the condensate spins
in both coupling regimes. Increasing the condensate separation to $\ell_c$
reduces this correlation to 0.09, confirming that condensate spins then become
uncoupled. It is important to note that we observe partial phase coherence
between the condensates in both AF and F regimes, and the condensates are at
equal energies (see SI.~\ref{si:phase}).

\begin{figure*} \centering \includegraphics[width=0.8 \textwidth]{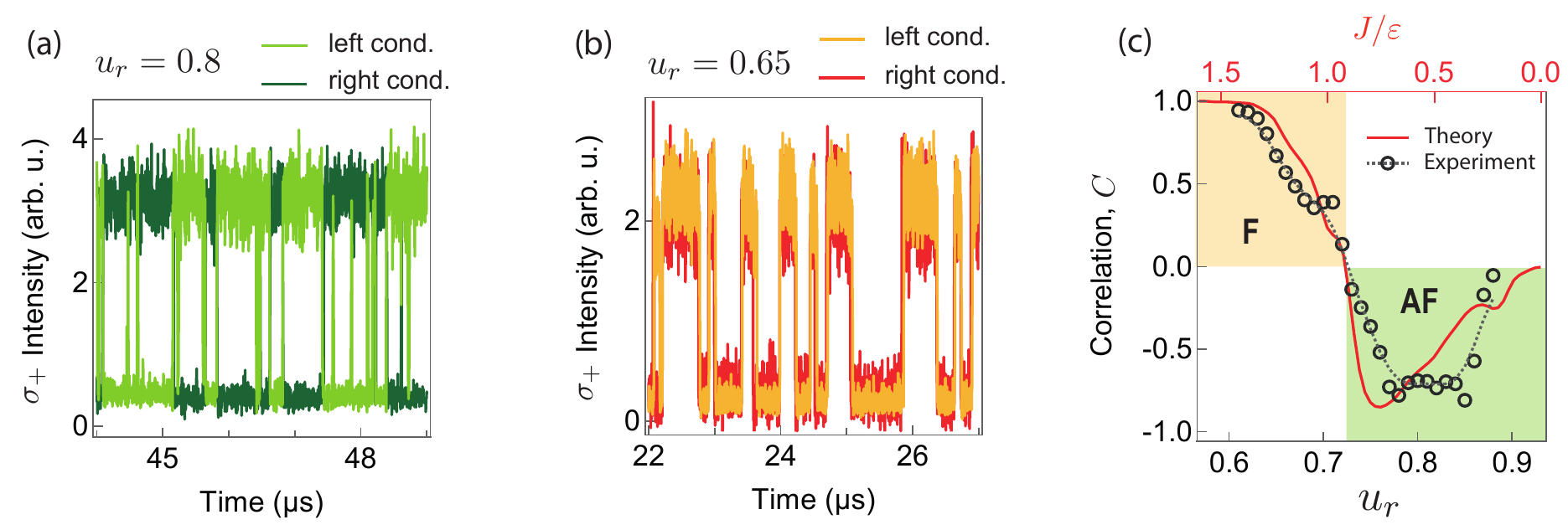}
    \caption{(a,b) Snapshot in time of left
        and right condensate $\sigma_+$ intensities for (a) $u_r$=0.80
        and (b) $u_r$=0.65, showing spontaneous correlated flipping of both
        condensates spins in (a) AF and (b) F regimes. (c) Magnetic phase
        diagram, with measured correlation $C$ between magnetization of condensates 1,2 as a function of
        barrier $u_r$ (circles, dotted line is guide for the
        eye).  Simulations give the correlation $C$ as a function of
        $J/\varepsilon$ (red line).}
    \label{fig:Chaotic} \end{figure*}

To accurately map the magnetic phase diagram of the system 
the influence of barrier on condensate spin correlation is investigated. Instead of
changing the separation of the condensates by changing the trap geometry, we
vary their barrier potential. For this the intensity of the barrier pump spots
[orange, Fig.~\ref{fig:DoubleHex}(a)] is changed. This allows finer control over the 
coupling interaction than changing the separation of the
condensates,
which is discretized due to single-pixel shifting of the SLM. To better observe the correlations, we induce spin flips by increasing the spin noise via spatially broadening the pump spots, which increases the overlap of the condensates and the pump. This increases the spontaneous spin-flip rate of the condensates (here set to  $\sim$10 flips/$\mu
\mathrm{s}$, see also SI.~\ref{si:spinnoise}), allowing faster
and more reliable correlation measurements within the stability time of our setup. Since the energy blueshift from pump-injected excitons is proportional to the
pump intensity~\cite{askitopoulos_polariton_2013}, the ratio of the intensity of
barrier pump spots to the other spots, $u_r$, is a reliable measure of the relative barrier height. The
blueshift above the condensate energy at the saddle point of the barrier is $u_r U_0$, where $U_0 \simeq \SI{200}{\ueV}$ (see also SI.~\ref{si:phase}). Selecting the right-circularly polarized ($\sigma_+$) 
emission, we  spatially resolve the left and right condensates recording their intensities at each $u_r$ for \SI{2}{\milli\second}
using photomultipliers. A typical trace for $u_r=0.8$ [Fig.~\ref{fig:Chaotic}(a)] shows that the double condensates flip randomly
between the two AF states with a switching time limited by our measurement
resolution (\SI{\sim 5}{\nano\second}).  Reducing $u_r$ to 0.65
shows now flipping between two F-states [Fig.~\ref{fig:Chaotic}(b)].
For each barrier potential we record 20 traces (each lasting
\SI{75}{\micro\second}) and calculate the average correlation of the
condensates spins, plotting this as a function of $u_r$ [Fig.~\ref{fig:Chaotic}(c)]. We observe a clear transition from the ferromagnetic to
antiferromagnetic state at $u_r\simeq 0.72$. Increasing $u_r$
 further results in zero coupling.

The uncoupling of the condensates when increasing their separation $\ell/\ell_c
= 0.9 \rightarrow 1.0$ (while the intensity of the shared pump spots remains
constant) implies that the shared reservoir between two condensates does not
play a significant role here.   Our trapped condensates form with $\bar{k}$=0
($\Delta k = \SI{0.4}{\radian/\um}$), where the transverse-electric and
transverse-magnetic splitting vanishes~\cite{shelykh_semiconductor_2004}.  As a
result the optical spin-Hall effect~\cite{kavokin_optical_2005} is negligible in
trapped condensates, with spin torque rates much smaller than tunnelling rates
thus preserving spin during the Josephson process~\footnote{At $k =
    \SI{0.4}{\radian/\um}$ the half-wavelength of the spin precession observed in
    Ref.~\citenum{kammann_nonlinear_2012} is \SI{\sim375}{\um}, which is more
    than 2 orders of magnitude larger than the condensate separations we have here.}.
On the other hand, observation of phase coherence between the condensates
signifies that the spin coupling must be mediated by a coherent mechanism, as
described by our theory below.

Our description of above effects is based on the theory for a single trapped
condensate~\cite{ohadi_spontaneous_2015}, which is extended to include the
Josephson coupling~\cite{wouters_excitations_2007,shelykh_josephson_2008,read_josephson_2010}
between the two condensates. The order parameter for each exciton-polariton
condensate is a two-component complex vector
$\Psi_n=[\psi_{n+},\psi_{n-}]^\mathrm{T}$ and $\psi_{n+}$ and $\psi_{n-}$ are the spin-up and
spin-down wave functions. The components of the order parameter define the
measurable condensate pseudospin
$\mathbf{S}_n=(1/2)(\Psi_n^\dag\cdot\bm{\sigma}\cdot\Psi_n)$, and the normalized
spin vector $\mathbf{\hat{s}}_n=\mathbf{S}_n/S_n$, where $\sigma_{x,y,z}$ are
the Pauli matrices. The order parameters evolve according to the driven
dissipative equation
\begin{align}
    \label{eq:main}
    \begin{split}
        \mi \frac{d\Psi_n}{dt}=&-\ihalf
        g(S_n)\Psi_n-\frac{\mi}{2}(\gamma-\mi\ve)\sigma_x\Psi_n
        \\&+\half\left[(\alpha_1+\alpha_2)S_n+(\alpha_1-\alpha_2)S_{nz}\sigma_z\right]\Psi_n
        \\&-\half J\Psi_{3-n}.
\end{split}
\end{align}
Here $g(S_n)=\Gamma-W+\eta S_n$ is the pumping-dissipation balance, $\Gamma$ is
the (average) dissipation rate, $W$ is the incoherent in-scattering (or
`harvest' rate), and $\eta$ captures the gain-saturation
term~\cite{keeling_spontaneous_2008}. This gain saturation depends on the total
occupation of the condensate (treated more generally in
Ref.~\cite{ohadi_spontaneous_2015}).  $X$ (horizontal) and $Y$ (vertical)
linearly-polarized single-polariton states are split in energy by
$\varepsilon$ and dissipation rate by $\gamma$. The repulsive interaction
constant for polaritons with the same spin is $\alpha_1$, and the
interaction constant for polaritons with opposite spins is $\alpha_2$. Finally, $J>0$ is the extrinsic spin-preserving Josephson coupling between the left and the right condensate, which we have introduced here to account for coherent coupling.

It is important to note the fundamental differences between the spin coupling
mechanism demonstrated here and that seen in closed systems such as atoms
trapped in optical lattices~\cite{struck_quantum_2011,greif_short-range_2013}.
In equilibrium spin systems such as those described by the Ising model, the
coupling is achieved via the minimization of the total energy. In the
driven-dissipative system described here, the minimization of energy does not
play a direct role since the system is out of equilibrium. Here the spin alignment is a
direct result of spin bifurcation correlation, which itself is a product of
pumping, dissipation and nonlinearity, and coherent exchange of particles
between the two condensates.

We perform dynamical simulations to calculate spin correlations. We calculate
the steady state of the coupled Eq.~\ref{eq:main} for \num{10000} realizations
with random initial conditions~\footnote{Similar to
    Ref.~\citenum{ohadi_spontaneous_2015}, the parameters used for 0D simulations
    are: $W=\SI{0.18}{\ps^{-1}}$; $\Gamma=\SI{.1}{\ps^{-1}}$;
    $\varepsilon=\SI{0.04}{\ps^{-1}}$; $\gamma=0.2\varepsilon$;
    $\alpha_1=\SI{0.01}{\ps^{-1}}$; $\alpha_2=-0.5\alpha_1$;
    $\eta=\SI{0.02}{\ps^{-1}}$} and plot the final circular polarization
correlation of the two condensates at different Josephson coupling rates
[Fig.~\ref{fig:Chaotic}(c), red line]. For small Josephson amplitudes there is
AF coupling of the condensate states. The stable AF configuration is
characterized by $\psi_{2+}=-\psi_{1-}$ and $\psi_{2-}=-\psi_{1+}$, so that
Eq.~\eqref{eq:main} is reduced to the two-component problem with renormalized
splitting between $X$- and $Y$-polarized states $\ve^\prime=\ve-J$.
Correspondingly, the AF state loses stability at $\ve=J$ when $\ve^\prime$
changes sign. The AF state is first converted to the limit cycle motion by the
Hopf bifurcation and then to pseudo-chaotic behavior with further increase of
$J$. Simultaneously, the stable F configuration with $\psi_{2+}=\psi_{1+}$ and
$\psi_{2-}=\psi_{1-}$ emerges for $J>\varepsilon$. This behavior is in good
agreement with the experiment: we observe the AF coupling at high barrier
heights where the tunneling rate between two condensates (Josephson coupling) is
small and same-site coupling between spins dominates
[Fig.~\ref{fig:DoubleHex}(a), white arrows]. On the other hand when the barrier
height is low, F-coupling is seen since the Josephson coupling dominates
[Fig.~\ref{fig:DoubleHex}(a), black arrows]~\footnote{We note that apart from
    most probable pure F and AF configuration, other attractors exist for the
    system~\eqref{eq:main} even away from the crossover region $J\approx\ve$. In
    particular, the F configuration for $J>\ve$ can be also achieved with the
    broken $1\leftrightarrow2$ symmetry.}. Each interacting pair of condensates
then forms a plaquette, and longer chains of condensates will form a bosonic
ladder~\cite{atala_observation_2014}, which will in future be useful to probe
with magnetic fields~\cite{van_der_zant_field-induced_1992}.
We note that paired condensates emit light at the same frequency, as it is also found experimentally.

The chaotic dynamics in the AF-F crossover region of the present system
is different from the chaos in resonantly excited pair of exciton-polariton
condensates studied before~\cite{solnyshkov_chaotic_2009}. The trapped
exciton-polariton condensates here are excited \emph{incoherently}, yielding
chaos in an autonomous dynamical system. The details of this chaos and its
effects on the emission from the trapped condensates will be studied elsewhere.

\begin{figure} \centering \includegraphics[width=0.46 \textwidth]{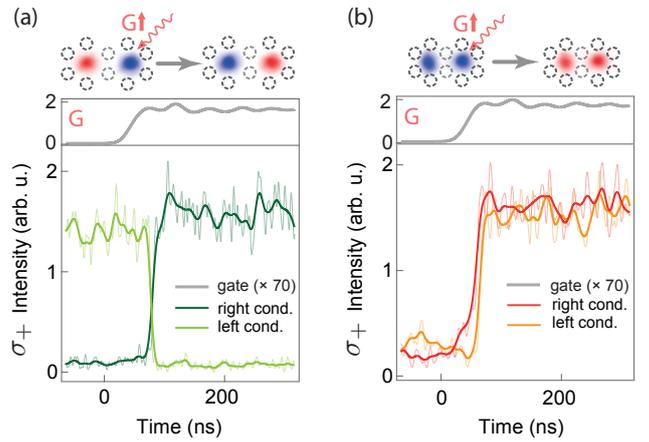}
    \caption{(a) Spin switching for AF-coupled condensates. The
        $\sigma_+$-polarized gate is applied on the right condensate (shown
        above). The condensate-pair switches from $\ket{\uparrow\downarrow}$ to
        $\ket{\downarrow\uparrow}$. (b) Spin switching for F-coupled
        condensates. The condensate-pair switches from
        $\ket{\downarrow\downarrow}$ to $\ket{\uparrow\uparrow}$. The top panels
        show the gate laser intensity profile G . The gate laser is turned on at
        $t=0$.} \label{fig:Switching} \end{figure}

Finally, we demonstrate the resonant switching of the coupled spin states experimentally. We
resonantly excite one of the condensates with a narrow linewidth CW diode laser,
which we refer to as the gate (G). The membrane microcavity allows resonant
excitation from the back side of the cavity without requiring filtering of the
laser backscatter. The gate laser as well as the pump can be switched on or off
by acousto-optic modulators (AOM) with a rise time of
\SI{\sim40}{\nano\second}.  We resonantly excite the right condensate
\SI{20}{\micro\second} after we turn on the pump laser. The gate is applied on
the right-hand condensate and is right-circularly polarized ($\sigma_+$).
Applying a cross-polarized gate switches the polarization state of
the condensate, as previously shown~\cite{ohadi_spontaneous_2015}. Here, we
observe that resonant switching of the spin of one condensate also switches the spin
of the other coupled condensate, both in the antiferromagnetic and ferromagnetic regimes (Fig.~\ref{fig:Switching}). The condensates remain in the
switched state after the gate laser is switched off, due to the bistable nature
of the spin states. We  note that the resolution-limited condensate switching time
reported here is an order of magnitude shorter than that of the gate, clearly showing that (as for single condensates~\cite{ohadi_spontaneous_2015}) the switching process is nonadiabatic. Here we are able to switch the coupled spin states by transiently injecting minority spins which are only 1\% of the condensates majority spin. Our theoretical description reproduces the fast spin flips observed in the experiment when noise is injected into Eq.~\ref{eq:main}. Our theory also shows that in both AF- and F-regimes, switching the spin state of condensate 1  also switches the spin of condensate 2, both remaining switched after the pulse is turned off (see SI.~\ref{si:resexc}).

In conclusion, we demonstrate a tunable spin coupling mechanism for trapped
polariton condensates. A transition from antiferromagnetic to
ferromagnetic coupling is seen as the potential barrier between the condensates
decreases. Our results correspond well to the interplay between spin-bifurcation and Josephson coupling in theory. 
Finally resonant switching of the
spin states in both ferromagnetic and antiferromagnetic regimes is shown.

\emph{Acknowledgments}---Authors acknowledge Andrew Ramsay, Dieter Jaksch and Tomi Johnson for fruitful discussions. This work was supported by EPSRC Grant No.
EP/G060649/1, EP/L027151/1, EU Grant No. INDEX
289968, ERC Grant No. LINASS 320503, the Leverhulme Trust Grant No.
VP1-2013-011, Spanish MEC (Grant No. MAT2008-01555), the Greek GSRT ARISTEIA
Apollo program and Fundación La Caixa, and Mexican CONACYT Grant No. 251808. FP acknowledges financial support through a Schr\"odinger Fellowship at the University of Oxford and NQIT project (EP/M013243/1). The data corresponding to the figures in this paper can be found at \url{https://www.repository.cam.ac.uk/handle/1810/253679}.

\bibliography{bib}


\clearpage
\pagebreak

\begin{center}
\textbf{\large Supplemental Information}
\end{center}
\setcounter{equation}{0}
\setcounter{figure}{0}
\setcounter{table}{0}
\setcounter{page}{1}
\makeatletter
\renewcommand{\theequation}{S\arabic{equation}}
\renewcommand{\thefigure}{S\arabic{figure}}
\renewcommand{\bibnumfmt}[1]{[S#1]}
\renewcommand{\citenumfont}[1]{S#1}
\renewcommand\thesection{\arabic{section}}
\graphicspath{{./SIfigs/}}

\section{Experimental methods}
\label{si:expmeth}
The cavity's top (bottom) distributed Bragg
reflector (DBR) is made of 32 (35) pairs of Al$_{0.15}$Ga$_{0.85}$As/AlAs layers
of \SI{57.2}{\nm}/\SI{65.4}{\nm}. Four sets of three \SI{10}{nm} GaAs quantum
wells (QW) separated by \SI{10}{nm} thick layers of Al$_{0.3}$Ga$_{0.7}$As are
placed at the maxima of the cavity light field.  The 5$\lambda$/2
(\SI{583}{\nm}) cavity is made of Al$_{0.3}$Ga$_{0.7}$As. The microcavity
sample is chemically etched from the substrate side to form \SI{300}{\um}
diameter membranes allowing optical access from the back of the sample for
resonant excitation. Note the phenomena reported in this paper do not depend on this etching, which is merely to conveniently explore the resonant gating regime. The sample shows condensation under nonresonant
excitation~\cite{tsotsis_lasing}. The CW pump is a single-mode Ti:Sapphire
laser tuned to the first Bragg mode $\SI{\sim 100}{\milli\electronvolt}$ above
the condensate energy, and is linearly polarized. In order to create short
time-scale realizations, the pump is amplitude-modulated by an AOM (rise time
$\simeq \SI{40}{\nano\second}$). A SLM is used to spatially pattern the pump
beam into a double hexagon using the MRAF
algorithm~\cite{MRAF}. The separation of the pump spots
can only be changed within the single-pixel shift limit of the SLM; however, the
intensity of the individual spots can be changed nearly arbitrarily. A 0.4 NA
objective is used for imaging the pattern onto the sample. The pulsed gate,
which is applied on the back side of the membrane, is a circularly-polarized CW
single-mode extended cavity diode laser, which is amplitude modulated using a
second AOM. A cooled CCD and a \SI{0.55}{\meter} spectrometer is used for imaging and energy resolving the photoluminescence (PL). PL is analyzed using a
quarter-waveplate and a Wollaston prism in front of the camera. Two
photomultiplier tubes (PMT) and a fast oscilloscope ($\SI{1}{\giga\hertz}$) are
used for time-resolving the polarization-resolved PL of the left and right
condensates. The CCD, the pump/gate laser and the oscilloscope are all synchronized together using arbitrary function generators ($\SI{20}{\mega\hertz}$).

\section{real-space}
\label{si:realspace}

\begin{figure}
\centering
\includegraphics[width=0.45 \textwidth]{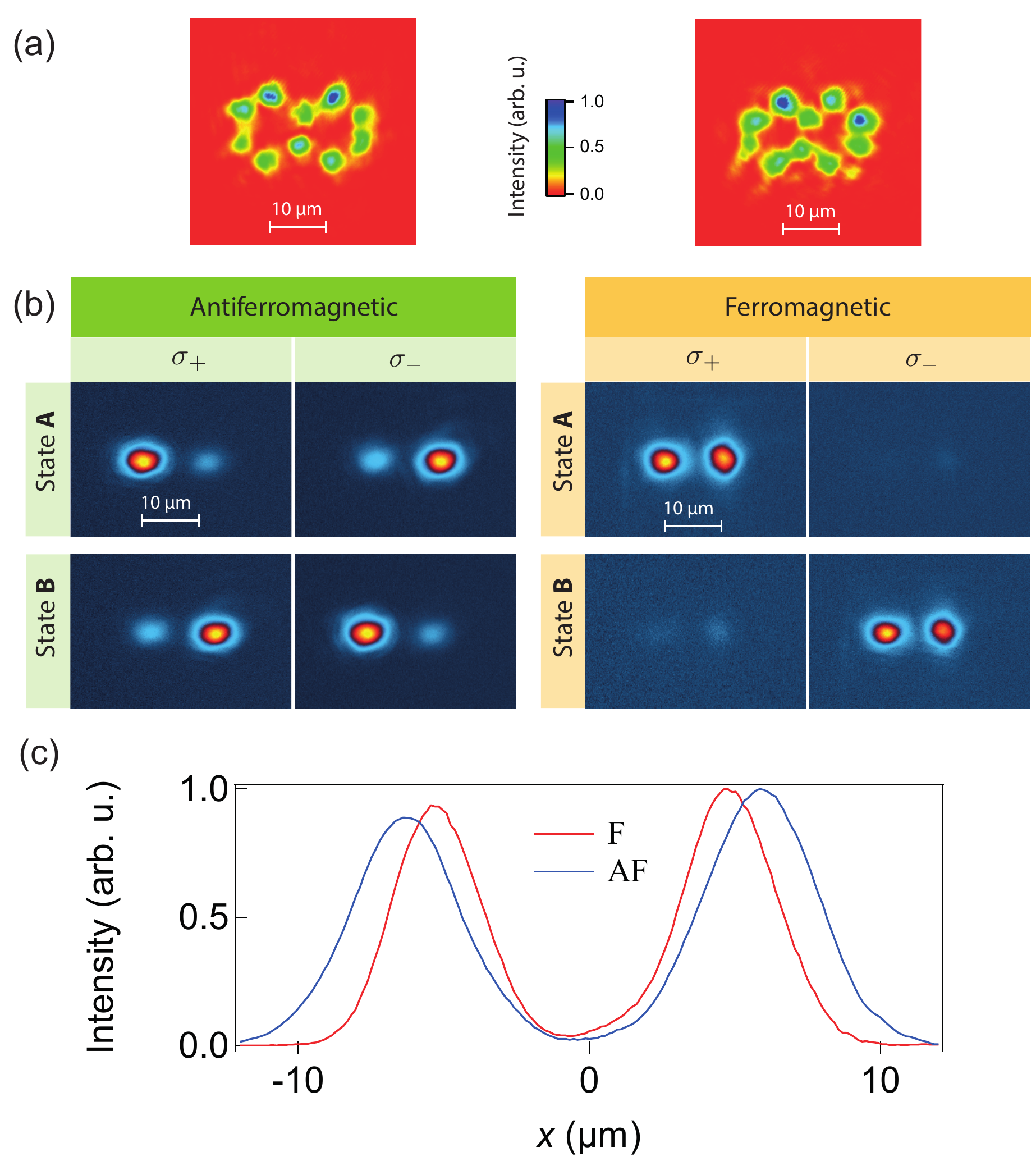}
\caption{(a) Pump intensity profiles in AF(left) and F-regimes (right) (b) Polarization-resolved real-space PL in AF (left) and F regimes (right). (c) Intensity profiles of the total emission in F and AF regimes.}
\label{fig:realspace}
\end{figure}

We can squeeze or stretch the trap without affecting the barrier pump spots [Fig.\ref{fig:realspace}(a)]. Using a Wollaston prism in front of the CCD, we measure both circular polarization components simultaneously for many realizations, as shown in Fig.~\ref{fig:realspace}. Here, each realization is a short-time exposure of the sample ($t\simeq \SI{200}{\micro\second}$).  The degree of circular polarization $M$ is given by $(i_{+}-i_{-})/(i_{+}+i_{-})$, where $i_{{\pm}}$ is the intensity image of the right- and left-circularly polarized emission. The emitted polarization is not exactly circular due to the spin bifurcation mechanism which shows the stable states have a small fraction of linear polarization~\cite{ohadi_spontaneous_2015_SI}. The spatial separation of the two condensates is extracted from the line profiles in Fig.~\ref{fig:realspace}(c).

\section{Phase and Energy}
\label{si:phase}
\begin{figure}[H]
\centering
\includegraphics[width=0.48 \textwidth]{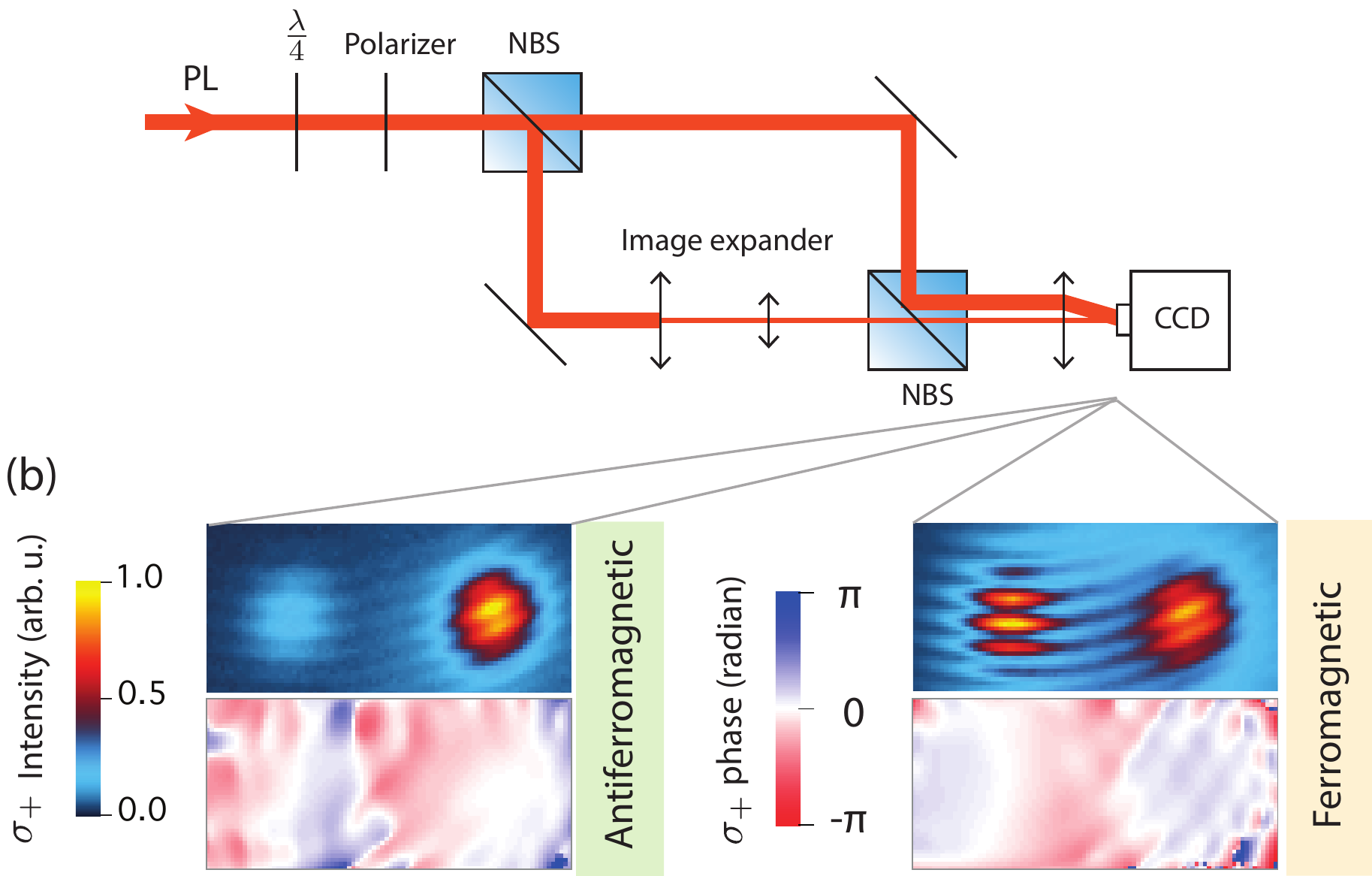}
\caption{(a) Michelson interferometer. (b) Interference pattern and extracted phase in AF (left) and F-regime (right).}
\label{fig:michelson}
\end{figure}

To experimentally study the phase coherence between the two condensates, a modified Michelson interferometer is used, as shown in Fig.~\ref{fig:michelson}. The circularly polarized PL is 5 times expanded in one arm and is superimposed onto that from the other arm. By doing this the PL of one condensate overlaps the PL from both condensates. The vertical displacement of the beams from the two arms on a final lens before the CCD results in fringes. Observation of fringes shows that we have partial phase coherence between the two condensates ($8\% <$ 1$^\text{st}$ order deg of coherence $<25\%$). The phase can be extracted by selecting the first-order diffraction of the Fourier transformed images and transforming it back into real-space images. We observe near zero phase difference in the circularly polarized basis of both AF and F regimes.

Phase fluctuations in exciton-polariton condensates are shown to be due to a
combination of condensate number fluctuations and the interaction of the
condensate with the excitonic reservoir~\cite{love_intrinsic_2008}. Both effects
perturb the phases of the two condensates independently because the two
condensates have separate reservoirs. On the other hand, the coherent exchange
of particles via Josephson coupling between the two condensates tries to lock
their phases. The mutual coherence is then set by the interplay between these
two effects. The phase coherence is zero in the case where the condensates are
completely uncoupled and gradually increases as the Josephson interaction
increases.

\begin{figure}
\centering
\includegraphics[width=0.42 \textwidth]{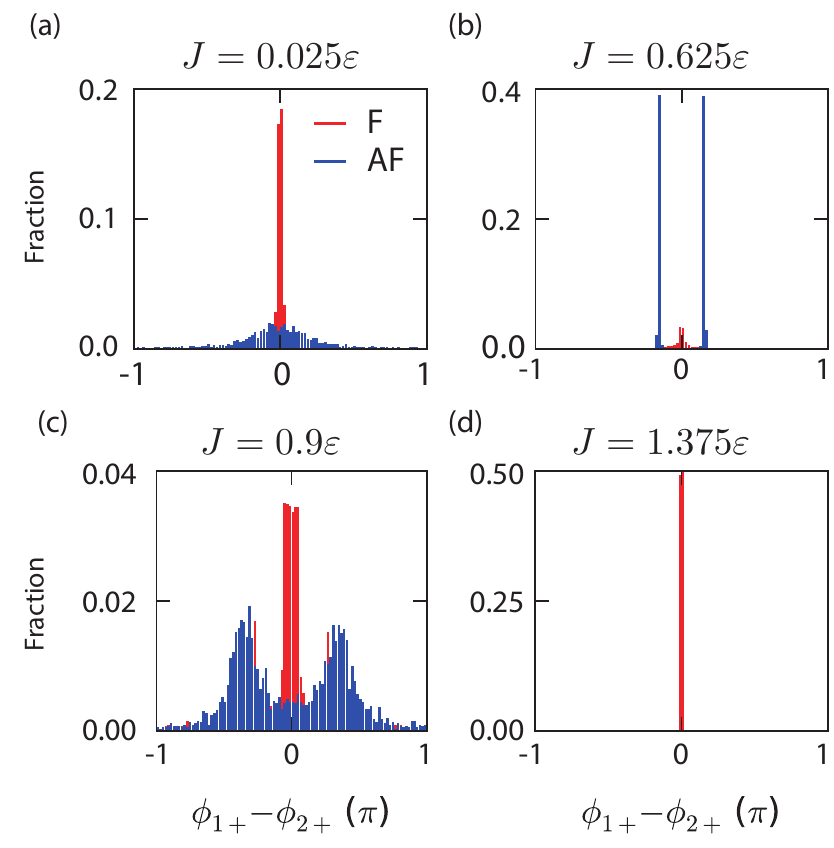}
\caption{Phase histogram for very small Josephson coupling (a), stable AF
    coupling (b), during the transition (c), and stable F-coupling (d).}
\label{fig:TheoryPhaseHist}
\end{figure}

Our dynamical simulation also predicts a nearly zero phase difference between
the same spin components in both F and AF regimes.
Fig.~\ref{fig:TheoryPhaseHist} shows a histogram of the phase difference between
same spin components of the left and right condensates ($\phi_{1+}-\phi_{2+}$)
as a function of the Josephson coupling rate $J$. For small Josephson couplings
($J=0.025\varepsilon$), F-states with zero phase difference appear, but their
fraction rapidly decreases as $J$ increases until two stable AF-states with
broken phase symmetry appear ($J=0.625\varepsilon$). At the transition point
($J=0.9\varepsilon$) the AF states destabilize and as $J$ increases further one
stable F-state appears ($J=1.375\varepsilon$).

\begin{figure}
\centering
\includegraphics[width=0.42 \textwidth]{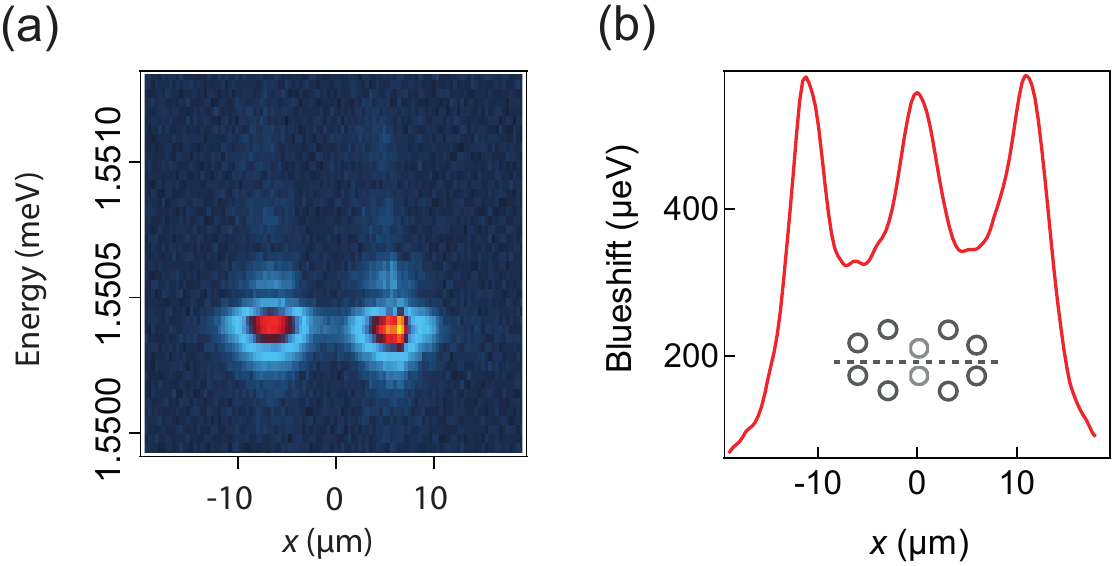}
\caption{(a) Energy profile of the condensates (b) Measured blueshift at a cross section of the hexagonal trap (shown by dotted line in inset) for $u_r=0.95$.}
\label{fig:Potential}
\end{figure}

Observation of interference fringes suggest that the two condensates have equal
energies. This is confirmed by energy resolving the condensate emission. We
observe that the two condensates emit light at the same frequency.  In the
experiment the energies of the two condensates are close ($<$\SI{100}{\ueV})
even when they are uncoupled because experimentally the two traps are symmetric,
i.e., the pattern intensity is the same for both traps.  Then the
coherent coupling between the two traps pulls the emission frequencies together,
in a similar fashion to the injection locking of two laser oscillators~\cite{stover_locking_1966}.

By measuring the blueshift of polaritons across the saddle points of the double-hexagon trap, we can extract the potential profile, as shown in Fig.~\ref{fig:Potential}. The potential barrier between the two traps is a linear function of the middle pump spots intensity and is $\SI{\sim 200}{\micro\electronvolt}$ at $u_r=1$.

\section{Pump-induced spin flips} 
\label{si:spinnoise}

\begin{figure}
\centering
\includegraphics[width=0.25\textwidth]{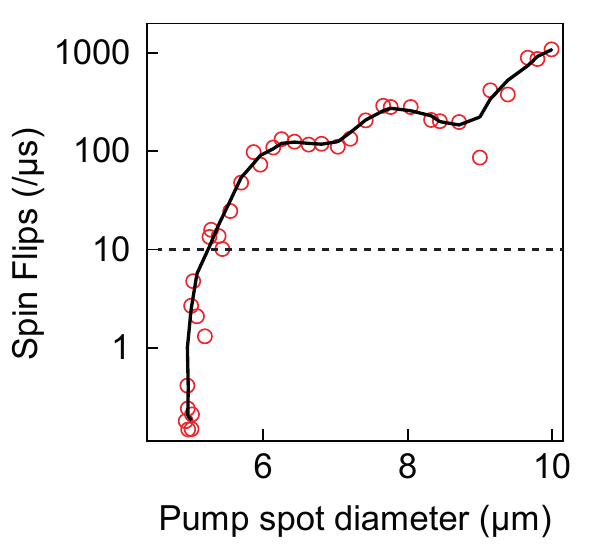}
\caption{Spin-flip rate as a function of pump spot diameter. The dotted line
    marks the rate where the correlated spin oscillations were measured. Black
    line is a guide for the eye.}
\label{fig:pumpSpotDiameter}
\end{figure}

In previous
work~\cite{ohadi_spontaneous_2015_SI}, we demonstrated how the interaction of the
condensate with pump-induced reservoir results in the reduction of the average condensate magnetization $M$. In that work we
noted: \begin{quote}``In a similar fashion to thermal noise, an overlapping reservoir can
also induce spin noise in the condensate. The spin noise causes condensate spin
flips, which result in the reduction of the time-averaged circular polarization.
Theoretically, this can be studied by introducing a noise term similar to Eq.
(9), but instead of depending on temperature, the noise intensity depends on the
overlapping reservoir density.''\end{quote}
Here, we quantify the influence of the reservoir on the spin-flip rate, as shown
in Fig.~\ref{fig:pumpSpotDiameter}. We measure the
rate of condensate spin-flips over $\SI{1}{\milli\second}$ as a function of pump
spot diameter. We observe that the condensate spin-flip rate increases
exponentially as the overlap between the reservoir and condensate increases and
plateaus, most probably due to finite resolution of our detection system. To
measure the spin correlations as a function of trap barrier, we broaden the pump
spots to $\sim \SI{10}{flips/\us}$, marked by a dotted line in
Fig.~\ref{fig:pumpSpotDiameter}.  

\section{resonant excitation and spin noise}
\label{si:resexc}

\begin{figure}[H]
\centering
\includegraphics[width=0.48 \textwidth]{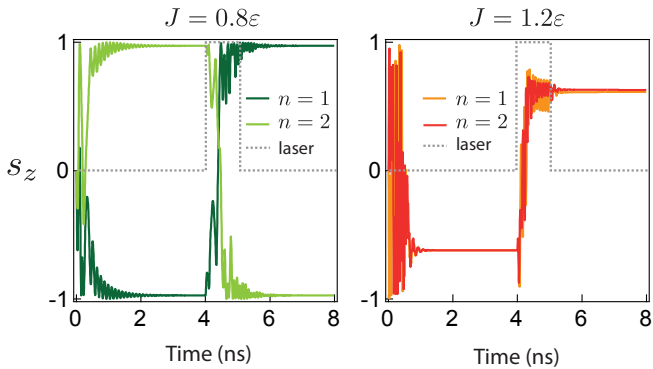}
\caption{The time evolution of two
        condensates for the case when $J=0.8 \varepsilon$ (b) and $J=1.2\varepsilon$ (c) is shown.
        The condensate-pairs switch their spin states when a gate pulse with an
        opposite spin is applied on condensate 1. The gate pulse is a spin-up
        square pulse that is turned on at $t=\SI{4}{\nano\second}$ and lasts for
        \SI{1}{\nano\second}. Parameters as in text with $\vert F \vert/\sqrt{N}=7.5\times10^{-3}$, where $N$ is the occupation of the condensates with $\omega_g$ at the condensate energy.}
\label{fig:TheorySwitching}
\end{figure}

\begin{figure}[H]
\centering
\includegraphics[width=0.42 \textwidth]{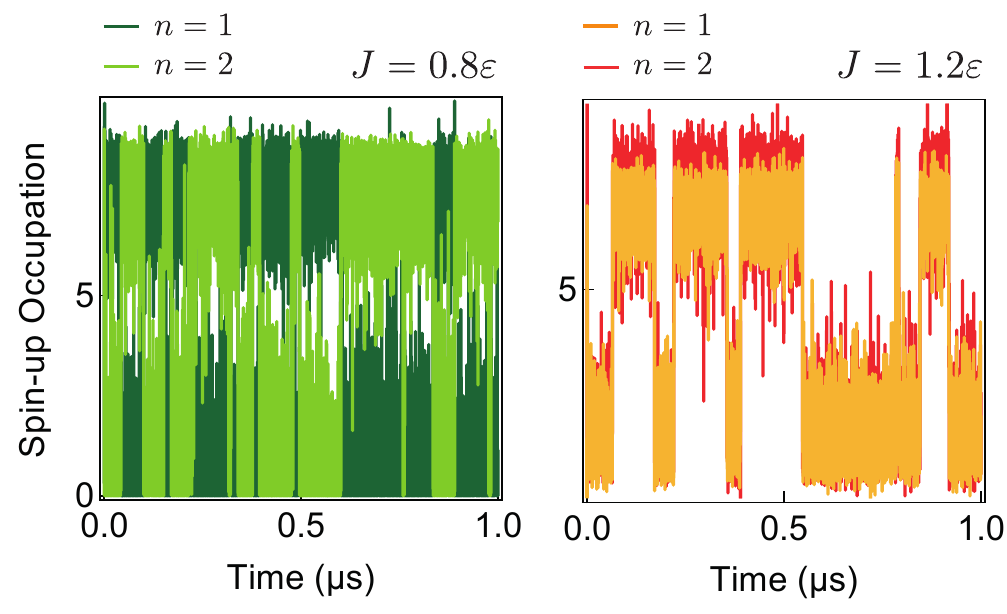}
\caption{Simulated noise-induced spin flips for AF- ($J=0.8\varepsilon$) and F-regimes
    ($J=1.2\varepsilon$).}
\label{fig:TheoryNoiseSwitching}
\end{figure}

To investigate resonant excitation and spin noise we modify our driven dissipative equation to
\begin{align}
    \label{eq:mainSI}
    \begin{split}
        \mi \frac{d\Psi_n}{dt}=&-\ihalf
        g(S_n)\Psi_n-\frac{\mi}{2}(\gamma-\mi\ve)\sigma_x\Psi_n
        \\&+\half\left[(\alpha_1+\alpha_2)S_n+(\alpha_1-\alpha_2)S_{nz}\sigma_z\right]\Psi_n
        \\&-\half J\Psi_{3-n} \qquad \text{(Josephson coupling)}
        \\&+F_n(\omega_g,t) \quad\;\;\; \text{(resonant excitation)}
        \\&+f_n(t). \qquad\;\quad \text{(spin noise)}
\end{split}
\end{align}
Here, $F$ is a two-component function representing resonant excitation at frequency
$\omega_g$.
To demonstrate spin-switching within our theoretical framework, we set the resonant
gate excitation $F$ to a right-circularly polarized ($\sigma_+$) pulse which is
applied to
condensate 1 at $t=\SI{4}{\nano\second}$ for a duration of \SI{1}{\nano\second}.
In both the AF-regime and the F-regime, when the spin state of condensate 1 is switched, condensate 2 also switches spin, as shown in Fig.~\ref{fig:TheorySwitching}. Moreover, both condensates stay in the switched state after the pulse is turned off.

To demonstrate noise-induced spin flips a spin noise function $f_n$ is added to Eq.~\ref{eq:mainSI}.  Here, $f_n(t)=[f_{n+}, f_{n-}]^T$ 
where $f_{\sigma}(t)$ with $\sigma=\pm1$ is a realization of Gaussian
random processes with zero mean $\langle f_{n\sigma}(t)\rangle=0$ and $\delta$-like
two-point correlation function:
\begin{align}\label{eq:noise}
 \langle f_{n\sigma}(t) f_{n\sigma'}(t')\rangle &=0, \\
 \langle f_{n\sigma}(t) f^*_{n\sigma'}(t')\rangle &=2 D\delta_{n,n'}\delta_{\sigma,\sigma'}\delta(t-t'),
\end{align}
where $D$ is the total noise intensity. Fig.~\ref{fig:TheoryNoiseSwitching} shows the influence of spin noise on the
coupled condensates. We observe that in both regimes, noise-induced spin flips
in one condensate switch the spin of the neighboring condensate, in
agreement with the experiment.

\end{document}